\documentclass[conference,a4paper]{IEEEtran}

\usepackage{algorithm} 
\usepackage{algpseudocode} 

\usepackage{amssymb} 
\usepackage{graphicx} 

\usepackage{array} 
\usepackage{todonotes}

\usepackage{graphicx}
\usepackage{caption}
\usepackage{subcaption}

\usepackage{cite}

\newcommand\copyrighttext{%
  \footnotesize \textcopyright 2019 IEEE. Personal use of this material is permitted.
  Permission from IEEE must be obtained for all other uses, in any current or future
  media, including reprinting/republishing this material for advertising or promotional
  purposes, creating new collective works, for resale or redistribution to servers or
  lists, or reuse of any copyrighted component of this work in other works.}
\newcommand\copyrightnotice{%
\begin{tikzpicture}[remember picture,overlay]
\node[anchor=south,yshift=34pt] at (current page.south) {\fbox{\parbox{\dimexpr\textwidth-\fboxsep-\fboxrule\relax}{\copyrighttext}}};
\end{tikzpicture}%
}

\begin{document}

\title{FlowDyn: Towards a Dynamic Flowlet Gap Detection using Programmable Data Planes}

\author{
\IEEEauthorblockN{Cristian Hernandez Benet}
\IEEEauthorblockA{Computer Science Department \\
Karlstad University \\ 65188 Karlstad, Sweden \\
Email: cristian.hernandez-benet@kau.se}
\and
\IEEEauthorblockN{Andreas J. Kassler}
\IEEEauthorblockA{Computer Science Department \\
Karlstad University \\ 65188 Karlstad, Sweden
\\
Email: andreas.kassler@kau.se}}



\maketitle
\copyrightnotice

\IEEEpubidadjcol

\begin{abstract}
Data center networks offer multiple disjoint paths between Top-of-Rack (ToR) switches to connect server racks providing large bisection bandwidth. An effective load-balancing
mechanism is required in order to fully utilize the available capacity of the multiple paths. While packet-based load-balancing can achieve high utilization, it suffers from reordering. Flow-based load-balancing such as equal-cost multipath
routing (ECMP) spreads traffic uniformly across multiple
paths leading to frequent hash collisions and suboptimal performance. Finally, flowlet based load-balancing such as CONGA or HULA splits flows into smaller units, which are  sent on different paths. Most flowlet based load-balancing schemes depend on a proper static setting of the flowlet gap, which decides when new flowlets are detected. While a too small gap may lead to reordering, a too large gap results in missed load-balancing opportunities. In this paper, we propose FlowDyn, which dynamically adapts the flowlet gap to increase the efficiency of the load-balancing schemes while avoiding the reordering problem. Using programmable data planes, FlowDyn uses active probes together with telemetry information to track path latency between different ToR switches. FlowDyn calculates dynamically a suitable flowlet gap that can be used for flowlet based load-balancing mechanism. We evaluate FlowDyn extensively
in simulation, showing that it achieves 3.19 times smaller flow completion time at 10\% load and 1.16x at 90\% load.
\end{abstract}


\IEEEpeerreviewmaketitle

\section{Introduction}
\label{intro}
Data center networks 
provide a large bisection bandwidth to cope with capacity demands and latency requirements of distributed applications. Typically, many parallel paths are available between any Top-of-Rack (ToR) pairs, which requires effective load-balancing mechanisms in order to fully utilize the available capacity. Because of the diversity of services hosted in a data center, there are diverse types of traffic that need to be served concurrently. While some applications require low latency, others require large throughput for e.g. file transfer \cite{curtis2011mahout}. For effective load-balancing it is important not only to distribute the load across all paths but to do it efficiently without compromising the performance of the transport application, for instance, by introducing re-ordering.

Many data center networks have adopted Equal-Cost Multi-Path (ECMP) \cite{hopps2000analysis} routing, which assigns each flow randomly to one of several paths. When two elephant flows collide on the same path, performance with ECMP is significantly degraded \cite{vanini2017let,dixit2013impact}. Several load-balancing schemes  have been proposed that use different granularity, including flow- \cite{al2010hedera}, flowlet- \cite{alizadeh2014conga} and  packet-based \cite{ghorbani2017drill} schemes. While packet-based load-balancing schemes can rapidly react to changes in the path utilization and balance the traffic more efficiently, they may lead to reordering that severely impacts the performance of transport protocols such as TCP. On the other hand, flow-based load-balancing schemes do not cause any packet reordering but result in poor performance when hash collisions occur. Therefore, numerous efforts have been directed towards flowlet-based load-balancing schemes such as HULA \cite{katta2016hula}, CONGA \cite{alizadeh2014conga}, LetFlow \cite{vanini2017let}, Flare \cite{sinha2004harnessing}, CLOVE \cite{katta2017clove}, which try to avoid reordering while achieving high performance at the same time.

Software Defined Networking (SDN) emerged as a prominent paradigm that overcomes the limitations of traditional network architecture. However, the lack of programmability of the forwarding plane has caused architectures such as Protocol Independent Switch Architecture (PISA) to emerge. Such programmable data plane architecture together with P4 programming language and compiler support \cite{bosshart2014p4} enables unprecedented flexibility and high-level abstractions for customized packet parsing, header definitions, logical sequence of the ingress and egress pipeline, etc. while being independent of data plane hardware. A P4 program can be compiled to different hardware targets, which allows  contributing to the innovation and emergence of new functionalities in different networking fields such as load-balancing \cite{katta2016hula}.

Flowlet-based load-balancing schemes split each flow into small flowlets, which are defined as a group of packets of a given flow that are separated in time by a large enough gap (denoted as flowlet gap) from the next flowlet in order to avoid reordering. When an inter-packet gap of more than the flowlet timeout is detected, a new flowlet is created that can be routed on a different path. Typically, a conservative static flowlet timeout of double the network round trip-time \cite{katta2017clove} is used. However, depending on server location, failures on the network, or a sudden increase in traffic demands, a static flowlet timeout may be either too small and consequently trigger unnecessary reordering or too large and reduce thus potential load-balancing opportunities.

In this paper, we propose FlowDyn, which uses programmable data planes to dynamically infer a good flowlet timeout, which can be used by flowlet-based load-balancing schemes. The main goal of FlowDyn is to increase the efficiency of the load-balancing schemes while avoiding the reordering problem. FlowDyn leverages periodic probes which are used by many load-balancing approaches, such as CLOVE  and HULA. 
FlowDyn tracks the path delay differences between ToR switches using programmable data planes (e.g. P4\cite{bosshart2014p4}). Throughout the paper, we use “flowlet gap” and “flowlet timeout” interchangeably. 
Finally, we evaluate FlowDyn in a packet-level simulator where we show its effectiveness compared to other well-known load-balancing schemes such as ECMP, HULA, and LetFlow. 

Our results show that a dynamic flowlet gap detection can achieve 3.19 times smaller average flow completion time at 10\% network load and 1.16x at 90\% network load.

The rest of the paper is structured as follows. In section \ref{background}, we briefly summarize background and related work. Section \ref{architecture} presents the design of FlowDyn. In Section \ref{eval}, we evaluate our approach using different load-balancing approaches in a simulated data center network. Section \ref{conclusion} concludes the paper and outlines future work.

\section{Background and Related Work}
\label{background}
Flowlet switching was introduced by Kandula et. al. \cite{sinha2004harnessing}, which proposes a new load-balancing method taking advantage of TCP's burstiness. The main idea is to temporally split a flow into smaller groups of flowlets, which are a burst of packets followed by an idle time (or flowlet gap). If the flowlet gap  is longer than the latency difference between two paths, the next group of packets can be sent on a different path without causing reordering. 
The flowlet concept has been used in different areas, including load-balancing \cite{vanini2017let,alizadeh2014conga,katta2016hula,katta2017clove}, congestion control   \cite{perry2017flowtune,shi2019adaptive}, and performance modelling \cite{guo2018set}.

\textbf{Congestion control:} Typically, TCP congestion control is packet based and the sender adjusts its rate based on incoming acknowledgements. \cite{perry2017flowtune} proposes to change this behavior to react on flowlet granularity. Flowlet based congestion control allows converging to an optimal rate allocation within just a few packets. On the other hand, \cite{shi2019adaptive} improves congestion control to make better decisions and thus improve the overall performance by better monitoring of congestion states and routing flowlets over least congested paths.

\textbf{Load-balancing:} The benefits of the flowlet concept have been applied to ECMP \cite{vanini2017let}. Different load-balancing schemes have been proposed exploiting the flowlet concept further. HULA \cite{katta2016hula} uses programmable data planes to track congestion states on all paths towards each ToR. Congestion information is obtained from probes that are sent through the network to collect hop-by-hop link utilization. Once the flowlet timeout expires, i.e., when the idle time between burst of packets is larger than a defined threshold, the next flowlet is routed along the least congested path. CLOVE uses ECMP and balances the load by manipulating packet-header fields at the hypervisor, which results in a different hash value and therefore forcing flowlets to be rerouted into a less congested path.

\textbf{Performance modeling:} 
 The performance impact of using flowlets has been studied in e.g. \cite{vanini2017let}. \cite{guo2018set} has tried to theoretically model the minimum timeout value by using a stationary Markov chain using strong assumptions. 
 
 Common to all work is that authors assume a static flowlet timeout, which has been configured beforehand to be large enough to avoid reordering. However, in a data center network, different server pairs communicate over network paths that span different hops. For example, when two servers are attached to the same Point-of-Delivery (POD), the traffic does not pass through the core and thus traverses fewer hops than other traffic that needs to pass over core switches. Also, congestion along the paths may impact the latency. Consequently, a too-large static flowlet timeout may not detect all load-balancing opportunities and lead to low network utilization. A too-small flowlet timeout may lead to reordering, which negatively impacts TCP congestion control. In contrast, FlowDyn uses programmable data planes to track latency differences between paths and dynamically adjust the flowlet timeout which leads to better load-balancing decisions.

\section{Flowdyn: Design}
\label{architecture}
\subsection{Overview}
The main idea of FlowDyn is to dynamically adjust  the flowlet timeout according to the topology and network conditions inside the data center network. FlowDyn tracks the latency difference among the multiple paths between the ToR switches that connect the server pairs. For each ToR pair, we calculate a flowlet timeout, so when the idle time of TCP connections is larger than the flowlet timeout, the switch can make a new decision to send the next flowlet on a different path.

\subsection{Introductory example}

\begin{figure}[!tbp]
  \begin{subfigure}[b]{0.225\textwidth}
    \includegraphics[width=\textwidth]{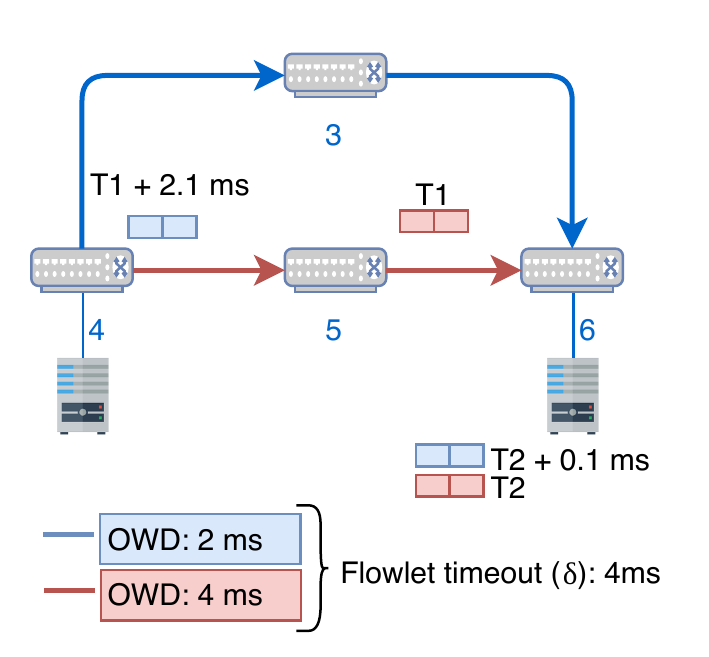}
    \caption{Missing load-balancing opportunity}
    \label{fig:flowlet_worst}
  \end{subfigure}
  \hfill
  \begin{subfigure}[b]{0.225\textwidth}
    \includegraphics[width=\textwidth]{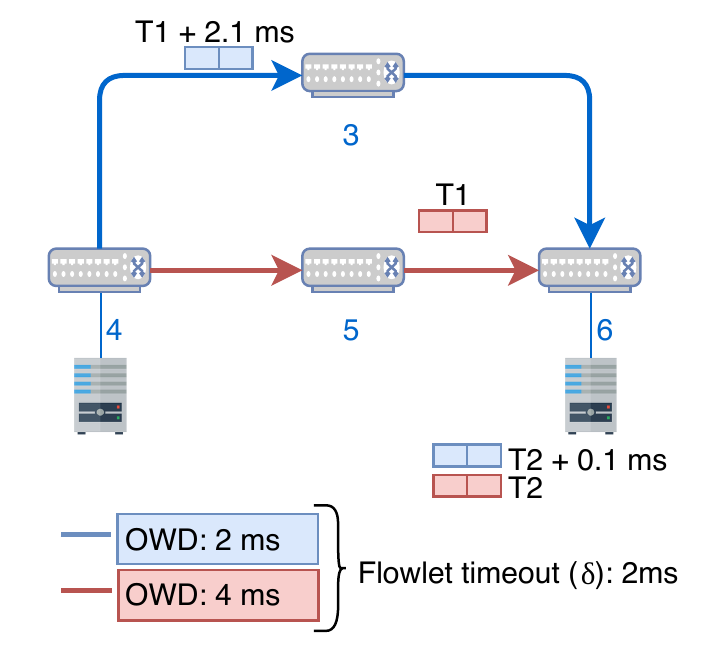}
    \caption{Ideal case flowlet timeout detection}
    \label{fig:flowlet_ideal}
  \end{subfigure}
  \caption{Flowlet timeout impact on load-balancing}
\end{figure}

The flowlet timeout plays a fundamental role in avoiding TCP reordering. A too-small timeout may lead to reordering if the new flowlet is sent on a new path that has lower latency. Conversely, a too large flowlet timeout might reduce the effectiveness of the load-balancing scheme by missing opportunities to create new flowlets that could be routed on less congested paths. 
An example of the latter is illustrated in Figure \ref{fig:flowlet_worst} where two different paths are available to route packets towards the server rack. Assume, that due to congestion each path has a different one-way delay (OWD), the blue path has 2ms and the red path 4ms. We can therefore calculate the minimum flowlet timeout, i.e. the minimum threshold to avoid reordering, as the latency difference between the two paths, which is 2ms in our example. The flowlet timeout is usually set statically using as a rule of thumb twice the round trip time (RTT)  \cite{katta2017clove},  or the maximum end-to-end latency \cite{alizadeh2014conga}. Following \cite{alizadeh2014conga}, we would set a timeout of 4ms based on the red path latency. Assume, two bursts of packets arrive at ToR 4, one seen at time T1 and another at T1 + 2.1ms. Assume, the first group of packets is sent on the red path. When the second group of packets arrives at T1 + 2.1ms, we do not create a new flowlet and keep forwarding it on the red path since the idle time between the two groups of packets is not larger than the flowlet timeout (4ms). Therefore, we have missed an opportunity to balance the load due to a too large flowlet timeout. 

Returning to the example illustrated in Figure \ref{fig:flowlet_worst}, FlowDyn idea is to calculate the flowlet timeout from the latency difference of the two paths. Therefore, we can reduce the flowlet timeout to 2ms and still ensure that there will be no reordering. Consequently, when the second group of packets arrives at  time T1 + 2.1ms, FlowDyn detects a new flowlet and make a new routing decision sending it over the blue path as shown in Figure \ref{fig:flowlet_ideal}. In this case, ToR 6 receives the first and second group of packets at time T2 and T2 + 0.1 ms, without causing reordering.

We now describe how we use programmable data planes to detect the different paths, how we track the latency difference among the fastest and the slowest path and how we calculate a suitable flowlet timeout in order to cope with microbursts that impact latency on small time scales. The ToRs perform the calculation of the flowlet timeout and are also responsible for monitoring the network and keeping updated the calculated flowlet timeout for each destination ToR. As a result, the intermediate nodes of the network do not have this information and it is necessary to provide it to avoid reordering and at the same time an optimal load balancing. This process of disseminating the flowlet timeout between ToR and the intermediate nodes of the network is explained in detail in section \ref{synchronization}.

\subsection{Path discovery}
In FlowDyn, each ToR periodically sends probes that are replicated through the network in order to discover all possible paths, similar to \cite{katta2016hula}. Probes are sent periodically by leaf ToRs at a certain frequency set through the control plane. The switches replicate the probes throughout the network  until the probe reaches a different ToR than the one that originated it. Then, the ToR processes the information stored in the probe (see later) and stops replicating it. The probes replicate both downstream and upstream within the same POD that contains the ToR that originated it, but only downstream when the probe reaches another POD. This approach prevents the appearance of undesirable loops. 
The probes allow FlowDyn not only to convey information about the network status such as the path latency but also discover new paths or network events such as link failures. The header fields of these probes can be extended to carry other types of network information, such as link utilization \cite{katta2016hula}.

\begin{figure}[!htbp]
  \centering
   \vspace{-1mm}
  \includegraphics[width=0.45\textwidth,keepaspectratio]{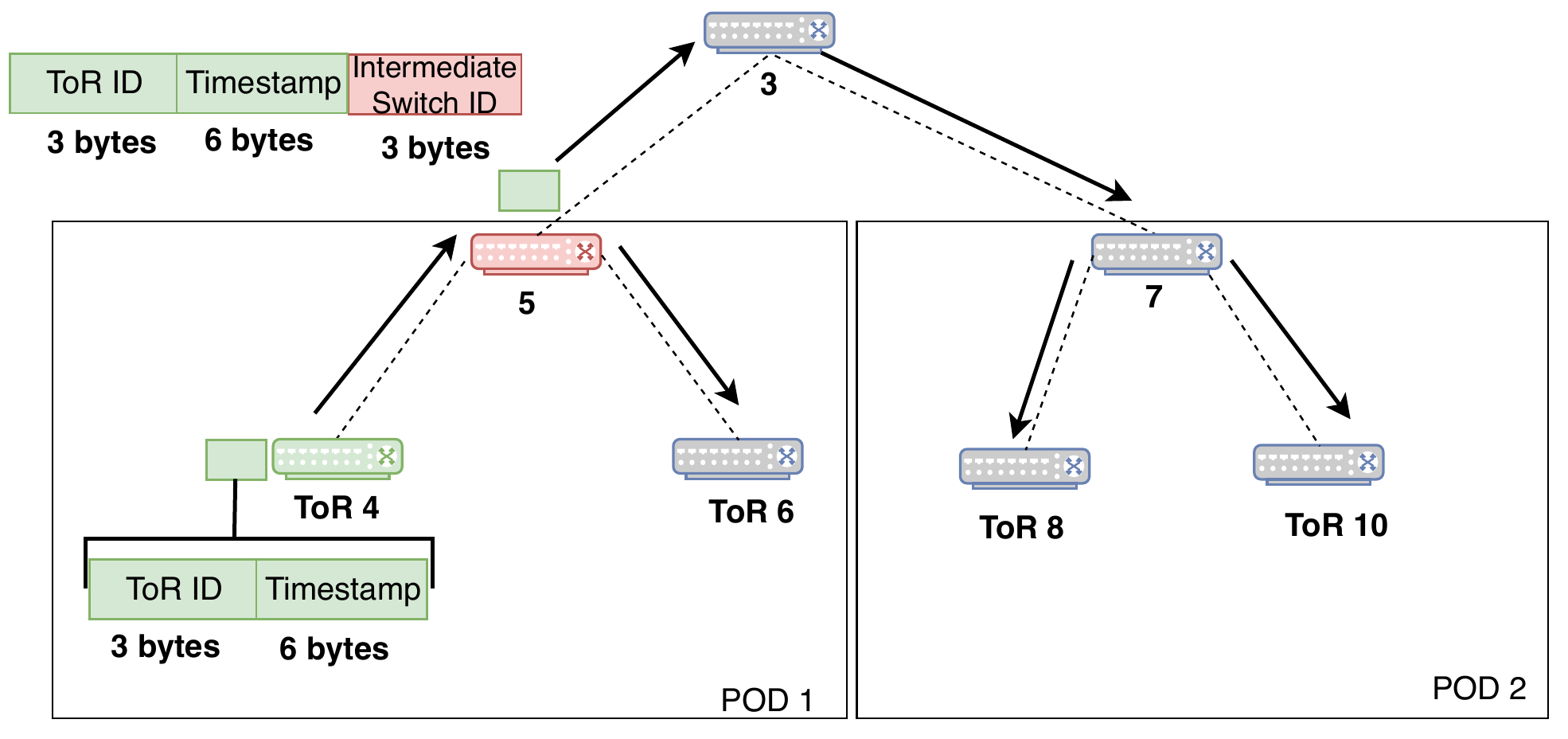}
  \caption{FlowDyn header and path discovery}
  \label{fig:flowdyn_header_dicovery}
  \vspace{-1mm}
\end{figure}

Each probe sent by the ToR contains at least the following information: the id of the ToR originating the probe and a timestamp field which contains the time when the probe was sent (step 1 in Figure \ref{fig:steps_flowdyn}). In addition, each intermediate switch that receives and replicates the probe, adds its own switch identification (switch id) to the probe, as illustrated in Figure \ref{fig:flowdyn_header_dicovery}. Therefore, the final size of the probe depends on the number of traversed hops.

Regardless of whether FlowDyn is used in combination with other methods to obtain other information from the network (e.g. probes in HULA obtain information about link congestion), FlowDyn utilizes 6 bytes for the timestamp information and 3 bytes for the identification of the ToR and intermediate switches. For example, for the topology  shown in Figure \ref{fig:topo}, where each ToR is at most 3 hops away from another ToR (without counting the last ToR), this will result in a total of 9 extra bytes to identify intermediate nodes, resulting in a total of 18 bytes for the FlowDyn header or a total of 19 bytes including link utilization for HULA load-balancing. The minimum packet size is 64 bytes, including the Ethernet, IP and FlowDyn header.

\subsection{Flowlet timeout detection}
ToRs transmit probes that are sent and replicated through the network to discover all paths until they reach a different ToR. The intermediate switches receive the probes originated from ToR, add their id to the header field of the probes and send them to the next-hop, as illustrated in step 1-2 of Figure \ref{fig:steps_flowdyn}. The next-hop is selected based on the chosen load-balancing scheme. FlowDyn calculates the flowlet timeout from the latency difference between two probes that travel through different paths. Since the flowlet timeout calculation is performed by the latency difference between probes originated from the same ToR, there is no need to have switch synchronization in the network. We use P4 and programmable data planes \cite{bosshart2014p4} to store in two registers arrays, \texttt{min\_remote table} and \texttt{max\_remote table}, the following information for each destination ToR: the ToR id where the probe originated from, the one-way delay (OWD) of the probe, the hash of the traversed path by the probe (denoted as H(P)) and the arrival time of the probe at ToR received. The arrival time of the probe is also stored in the register array since it serves as a timeout to detect when a table is not being updated due to the loss of probes. In such a case a sufficiently large flow timeout value is established. This can happen during large network congestion where probes might be dropped. 

The two register arrays discover and store the minimum and maximum latency between any pair of paths to the ToR destination. Therefore, whenever a probe with an OWD value lower than the one stored in the \texttt{min\_remote table} arrives at a ToR, the value of the table is updated with the information of the probe. In the same way, if ToR receives a latency greater than the one stored in the \texttt{max\_remote table}, we will update it with the information contained in the probe. In addition, we track the changes in the latency paths stored in the \texttt{min\_remote} and \texttt{max\_remote} tables by forcing to update the destination ToR entry when the ToR receives a probe which path is stored in any of the two tables. This allows detecting any latency rise of the path that is in the minimum registry or latency decrease of the maximum registry. Step 3 of Figure \ref{fig:steps_flowdyn} illustrates this process of storing the latency information in the min and max registers and performing the calculation of the flowlet timeout for each destination ToR from the delay difference of these two registers.

The ToR piggybacks the calculated flowlet timeout using any data or ACK packet sent towards the destination ToR (step 4 in Figure \ref{fig:steps_flowdyn}). When the ToR receives the piggybacked flowlet timeout value, it stores it in a register array called \texttt{local table} in step 5 of Figure \ref{fig:steps_flowdyn}. In this table, the ToR stores the destination ToR that has sent the probe, the timeout value towards that destination ToR, and the probe id. The probe id is used to detect old flowlet timeout values that may be received due to network delays. 

\subsection{Flowlet timeout - Intermediate Nodes} \label{synchronization}
ToRs obtain the flowlet timeout value through the piggybacked process. However, in topologies such as Fat-tree, due to the large number of hops between ToRs, it is necessary that intermediate switches also are configured with a proper flowlet timeout so as not to miss load-balancing opportunities.

This is achieved by reading the flowlet timeout field attached to the data packets in the piggybacked process. In this way, the intermediate nodes are informed about the flowlet timeout to apply for that destination ToR. This value is stored in the flowlet register, just like other information such as the current next-hop, age bit or timestamp \cite{alizadeh2014conga}. The ToR or last node before the destination server is responsible for removing this extra header with the information about the flowlet timeout.

\subsection{OWD calculation}
The probe sent periodically by a ToR is replicated through the network to discover the adjacent nodes, which allows discovering and updating all possible paths between one ToR and another.

Figure \ref{fig:steps_flowdyn} shows how the probe is replicated and sent to both switch 3 and switch 5. These two switches, in turn, add their own switch id and forward it to ToR 6. Consequently, ToR 6 receives two probes, one through switch 3 and the other through switch 5. When ToR 6 receives each of the probes, it reads and parses the timestamp of the probe and calculates the OWD as the difference between the current time and the probe timestamp. This OWD is stored in one of the two tables, \texttt{min\_remote table} or \texttt{max\_remote table} depending on the latency value obtained and the value of the tables.

Since the probe is sent from the same ToR 4, we can calculate the latency difference denoted as delta as $\Delta$ = $OWD_{maxlatency}$ - $OWD_{minlatency}$. Since the latency difference is calculated for each destination ToR, i.e., from the probes received by each ToR, and in very short time intervals, we can assume that there is no risk of clock drifting or any other synchronization problem since registers are constantly updated with the OWD when new probes arrive.

\subsection{Microburst Problem}
Microbursts are a burst of packets sent in a very short time period that can imply loss of packets, an increase in latency or congestion in the network during a short duration. \cite{zhang2017high} shows that 90\% of the burst data centers are of short duration less than 200$\mu$s.

This is especially important in our case for two reasons: (1) The probes are sent at a certain frequency to monitor the network, and (2) the flowlet timeout are piggybacked to the origin ToR. As probes are sent at a given  interval, they may not detect microbursts. Even when detecting microbursts, the path latency differences may arrive too late at the source ToR to react and change the flowlet timeout to match the queue built-up.

\begin{figure}[!htbp]
  \centering
   \vspace{-1mm}
  \includegraphics[width=0.45\textwidth,keepaspectratio]{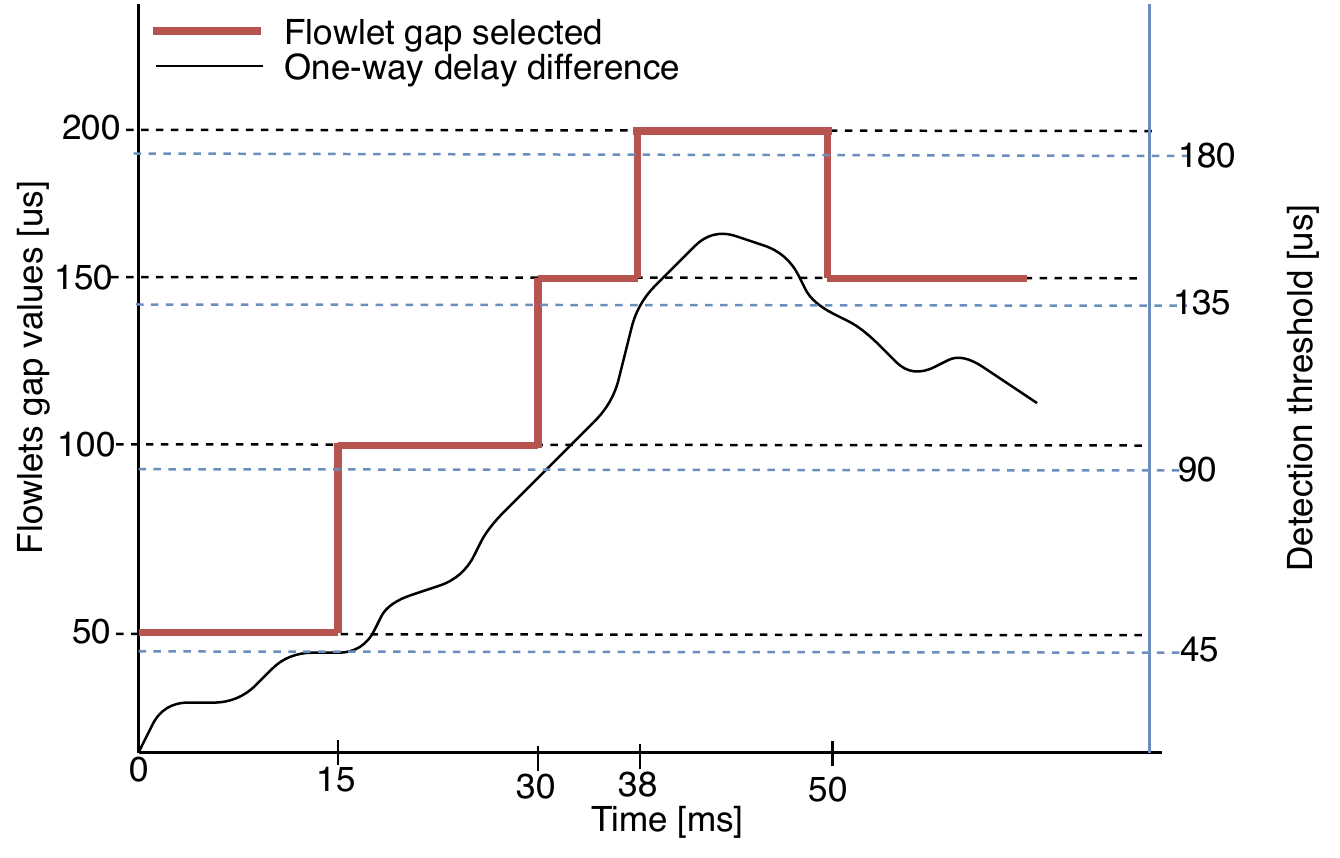}
  \caption{Flowlet gap selection}
  \label{fig:flowlet_thresholds}
  \vspace{-1mm}
\end{figure}

In FlowDyn we can detect one-way delay changes in time scales of the order of RTT, but microburst can occur at sub-RTT time periods. In order to cope with these abrupt latency variations produced by microbursts, we add safe margins to the flowlet timeout piggybacked instead of using the instantaneous flowlet timeout value. These margins are applied by a step function that allows our system to be robust against microbursts. An example is illustrated in Figure \ref{fig:flowlet_thresholds}, where the blue line represents the measured one-way delay differences between fastest and slowest path, which is piggybacked to the source ToR. The red line represents our step function where we have applied some margins to cope with microburst delay spikes. In this example, the step unit is in 50$\mu$s scale and the detection threshold is set to 90\% of the step function. When the one-way delay difference exceeds 90\% of the current step value (detection threshold), the selected flowlet gap jumps up to the next step (e.g. when one-way delay difference increases above 45$\mu$s, the selected flowlet gap is 100$\mu$s). Similarly, when the real flowlet timeout value is below the detection threshold, the flowlet gap selected jumps down to the prior value step (e.g. when one-way delay difference decreases below 135$\mu$s, the flowlet gap will be reduced from 200 to  150$\mu$s).

\subsection{Feasibility in P4}
To implement our approach in P4, we require both stateless and stateful operations. All network nodes need stateless operations for writing and reading the header fields of data and probe packets. For example, the intermediate switches in the network need to write their switch id in the probe header fields since this is necessary to identify the path and calculate the hash of the path. The ToRs need to perform stateful operations for recording and manipulating delay states.

\begin{figure}[!htbp]
  \centering
   \vspace{-1mm}
  \includegraphics[width=0.45\textwidth,keepaspectratio]{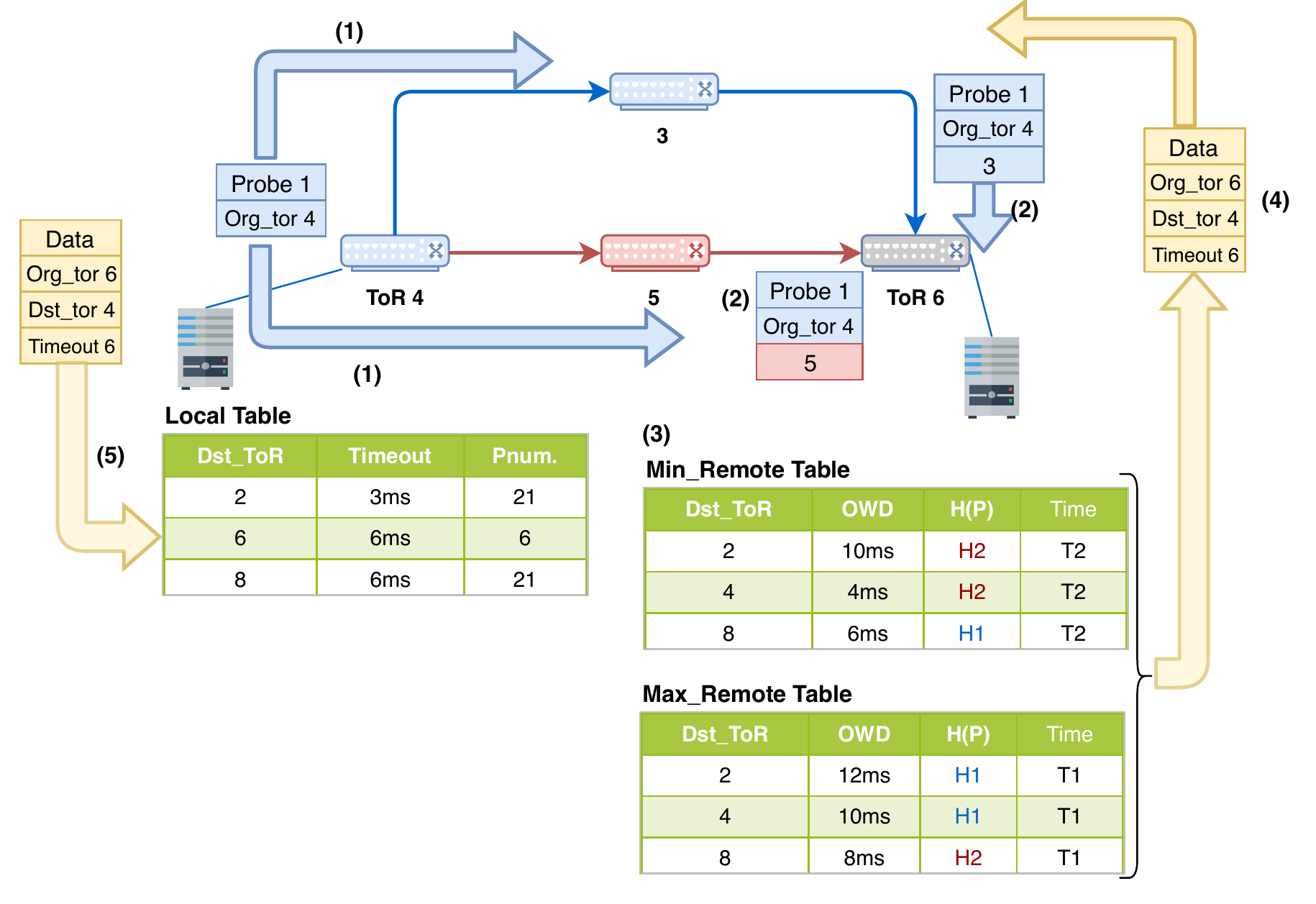}
  \caption{FlowDyn steps and registers}
  \label{fig:steps_flowdyn}
  \vspace{-1mm}
\end{figure}

\textbf{Parsing State:} 
FlowDyn requires processing and storing different information to calculate the flowlet timeout. In relation to processing, both the computation of the flowlet timeout and hash of path, are simple operations that can be done at line rate. In the case of the \texttt{min\_remote table} and \texttt{max\_remote} illustrated in Figure \ref{fig:steps_flowdyn}, ToRs need to store the illustrated information to compute the flowlet timeout of the remote destination ToR. For each entry in register array \texttt{min\_remote} and \texttt{max\_remote}, we need to store 13 bytes. Assuming 10K ToRs, the memory requirement is therefore around 130 KB for a single  table.

In addition, when the ToR receives the flowlet timeout through the piggybacked process, the flowlet timeout needs to be stored in a register array as shown in Figure \ref{fig:steps_flowdyn} (denoted as \texttt{local table}). In this table, we only need to store the information of the ToR destination, the flowlet timeout and the probe id, resulting in a total of 10 bytes per entry. Assuming 10K ToRs, we need 100KB of memory to store such information.

\textbf{Processing at ToR:} The ToRs are responsible for maintaining and manipulating the information of the three tables as illustrated in Figure \ref{fig:steps_flowdyn}. To perform the hash and identify the path that the probes have traversed, the stored information in the probe header is extracted. Therefore, this operation can be performed with minimal overhead \cite{bhamare2019intopt}.

\textbf{Processing at intermediate network nodes:} The intermediate nodes receive the flowlet timeout gap that is added by the ToRs to the data packets using an extra header. In this operation, intermediate nodes need to perform read and write operations. In addition, intermediate nodes need to add their switch id in the probes to facilitate the identification of the paths in the ToRs.

\section{Evaluation and Results}
\label{eval}
In this section, we evaluate the performance of FlowDyn by implementing and testing such approach using the NS2 packet-level simulator \cite{issariyakul2009introduction}. 

\begin{figure}[!htbp]
  \centering
   \vspace{-1mm}
  \includegraphics[width=0.45\textwidth,keepaspectratio]{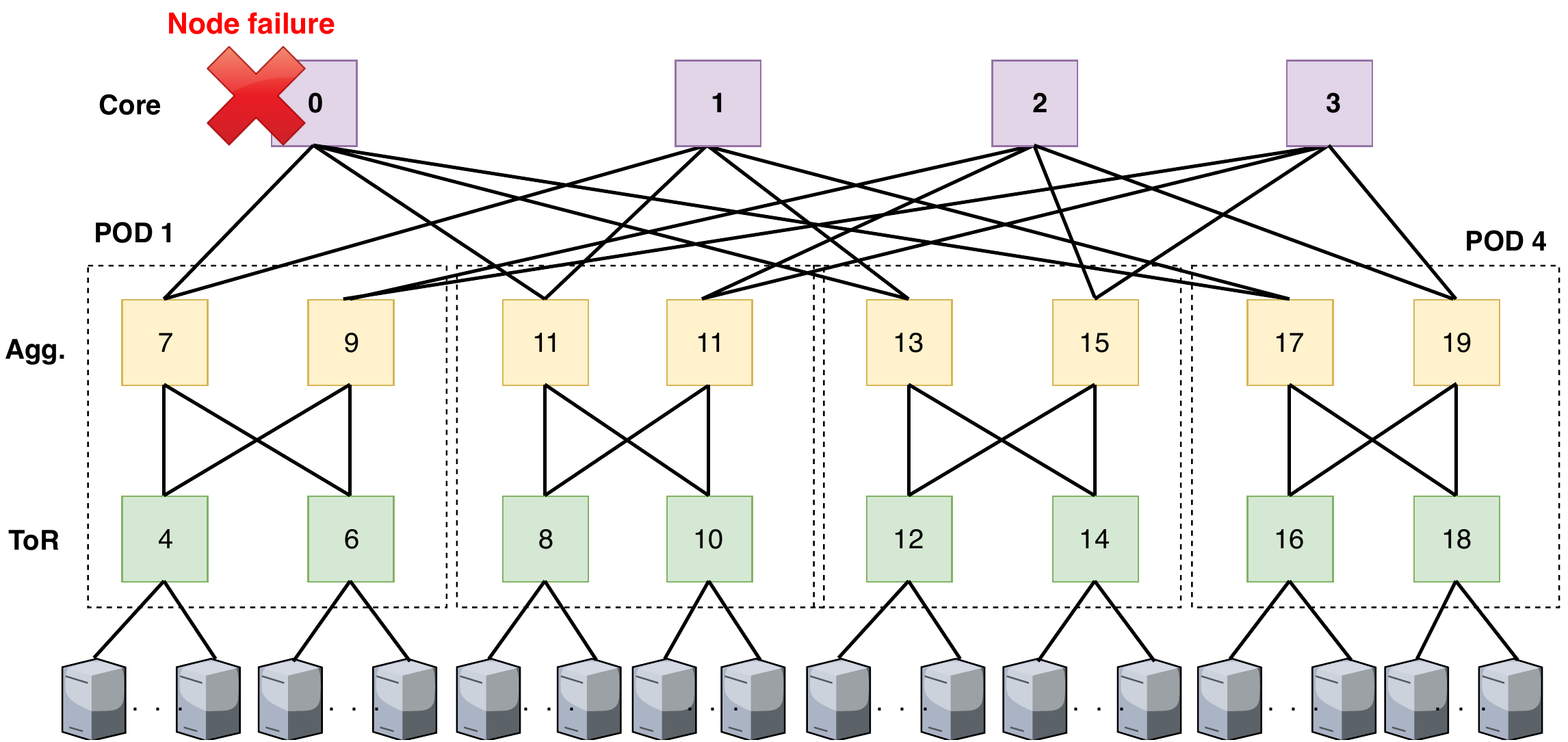}
  \caption{Evaluation Topology}
  \label{fig:topo}
  \vspace{-1mm}
\end{figure}

\textbf{Topology:} We use a 3-tier fat-tree topology with four core switches and four PODs containing each two aggregate and edge switches as illustrated in Figure \ref{fig:topo}. There are 8 hosts connected to each edge switch, thus having in total 64 hosts. All links have a capacity of 40Gbps except the links between the edge and hosts that have a capacity of 10Gbps. This leads to a total bisection bandwidth of 160Gbps avoiding an oversubscribed network. To simulate an asymmetric topology, we disable one of the core switches as illustrated in Figure \ref{fig:topo}.

\begin{figure*}[!htb]
   \begin{minipage}{0.325\textwidth}
     \centering
     \includegraphics[width=.9\linewidth]{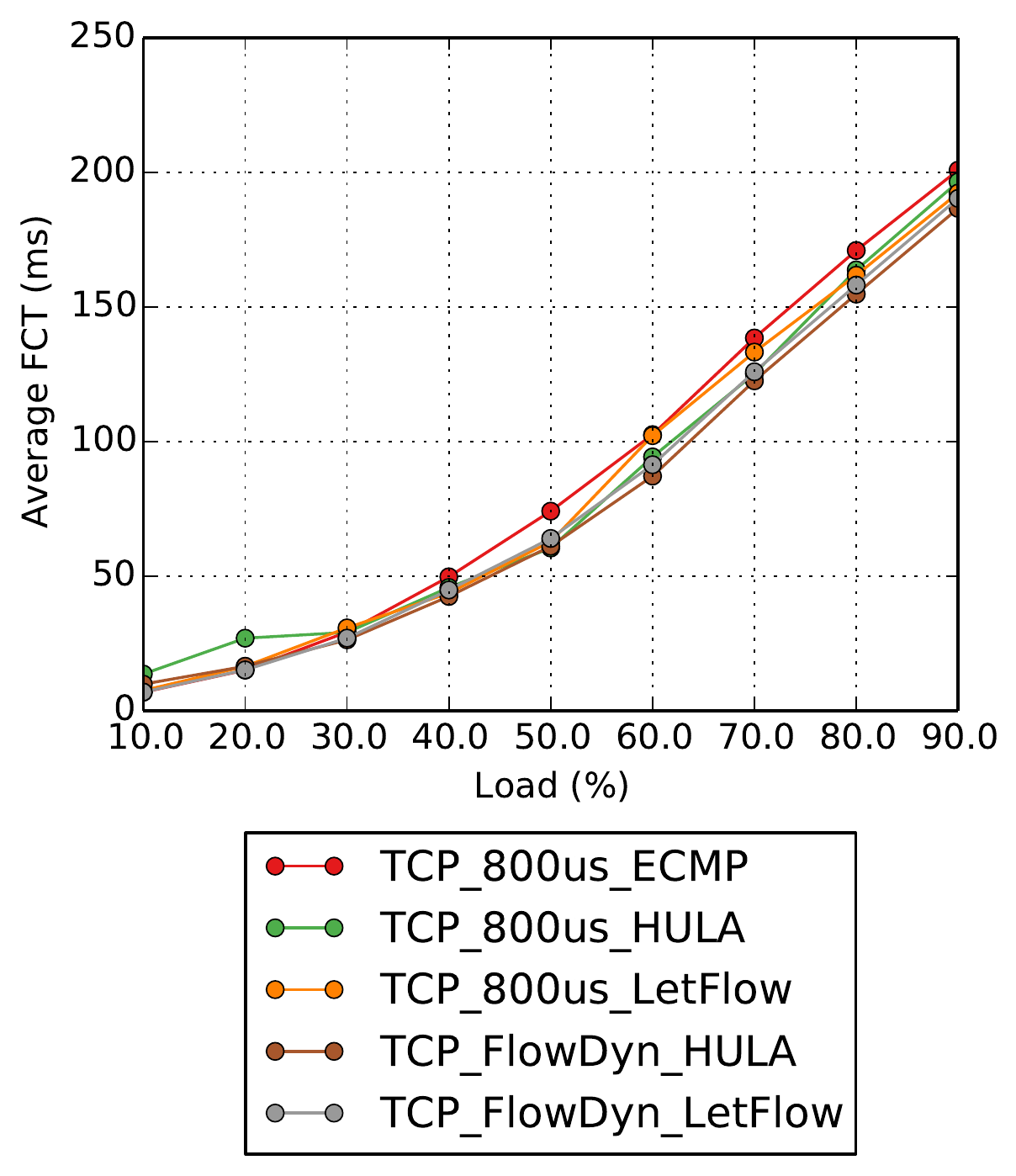}\par\caption{Average FCT using for web-search traffic} \label{fig:Average_fct_dctcp} 
   \end{minipage} \hfill
   \begin {minipage}{0.325\textwidth}
     \centering
     \includegraphics[width=.9\linewidth]{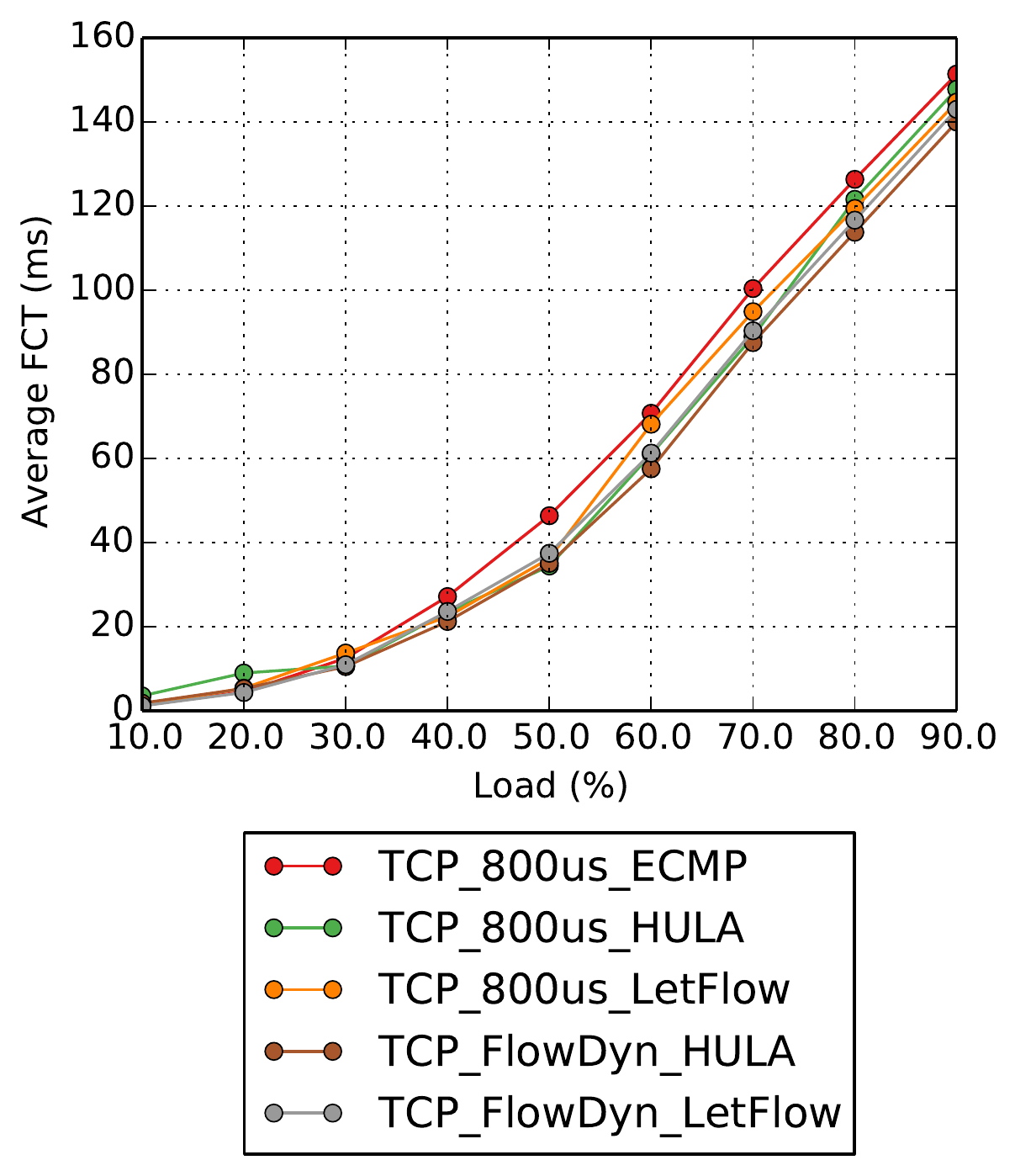}\par\caption{Average FCT for mice flows (\textless100KB), web-search} \label{fig:mice_websearch}
   \end{minipage} \hfill
     \begin {minipage}{0.325\textwidth}
     \centering
     \includegraphics[width=.9\linewidth]{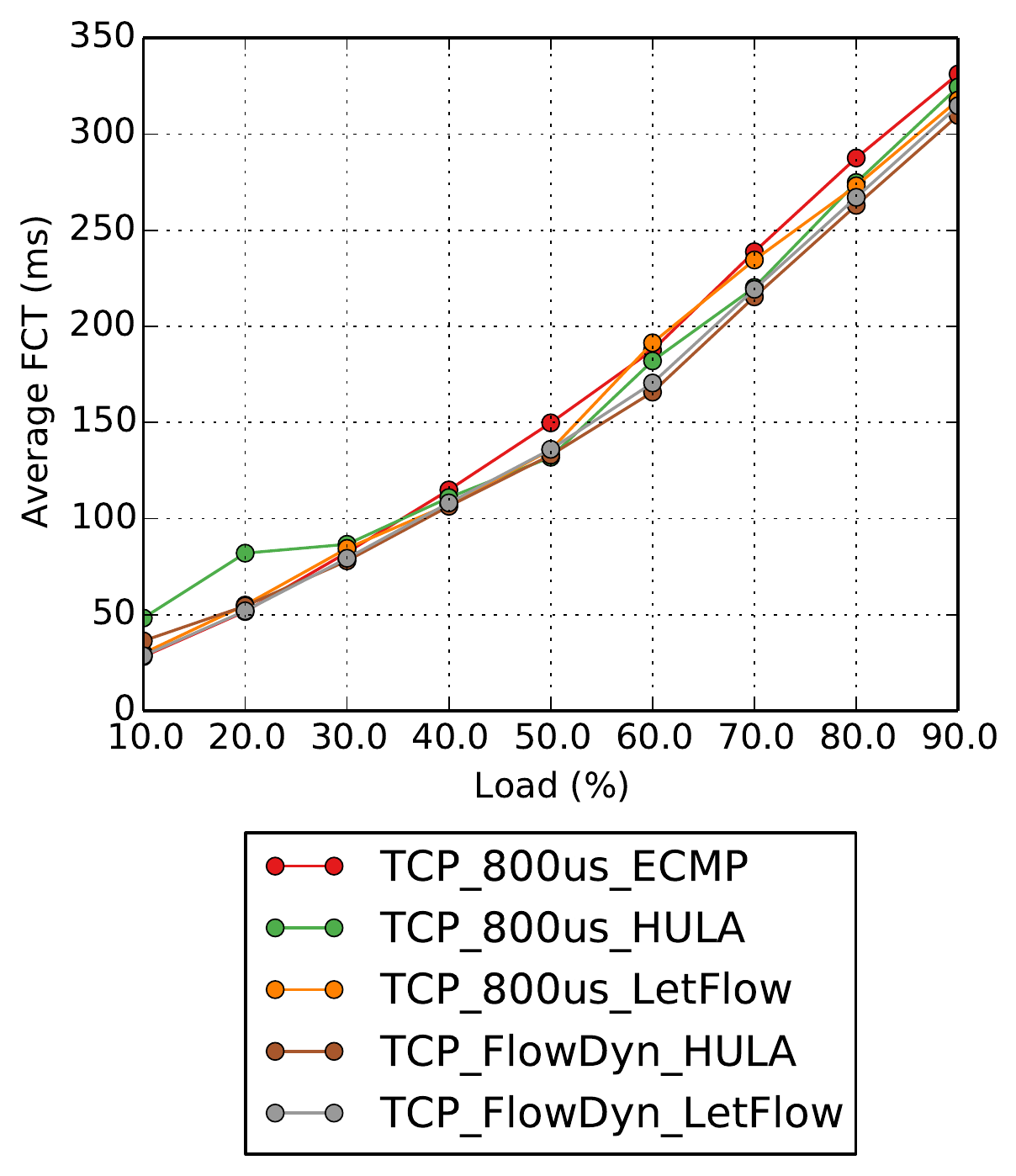}\par\caption{Average FCT using for elephant flows (\textgreater10MB) web-search traffic}  \label{fig:Average_elephant}
   \end{minipage}
\end{figure*}

\begin{figure*}[!htb]
   \begin{minipage}{0.325\textwidth}
     \centering
     \includegraphics[width=.9\linewidth]{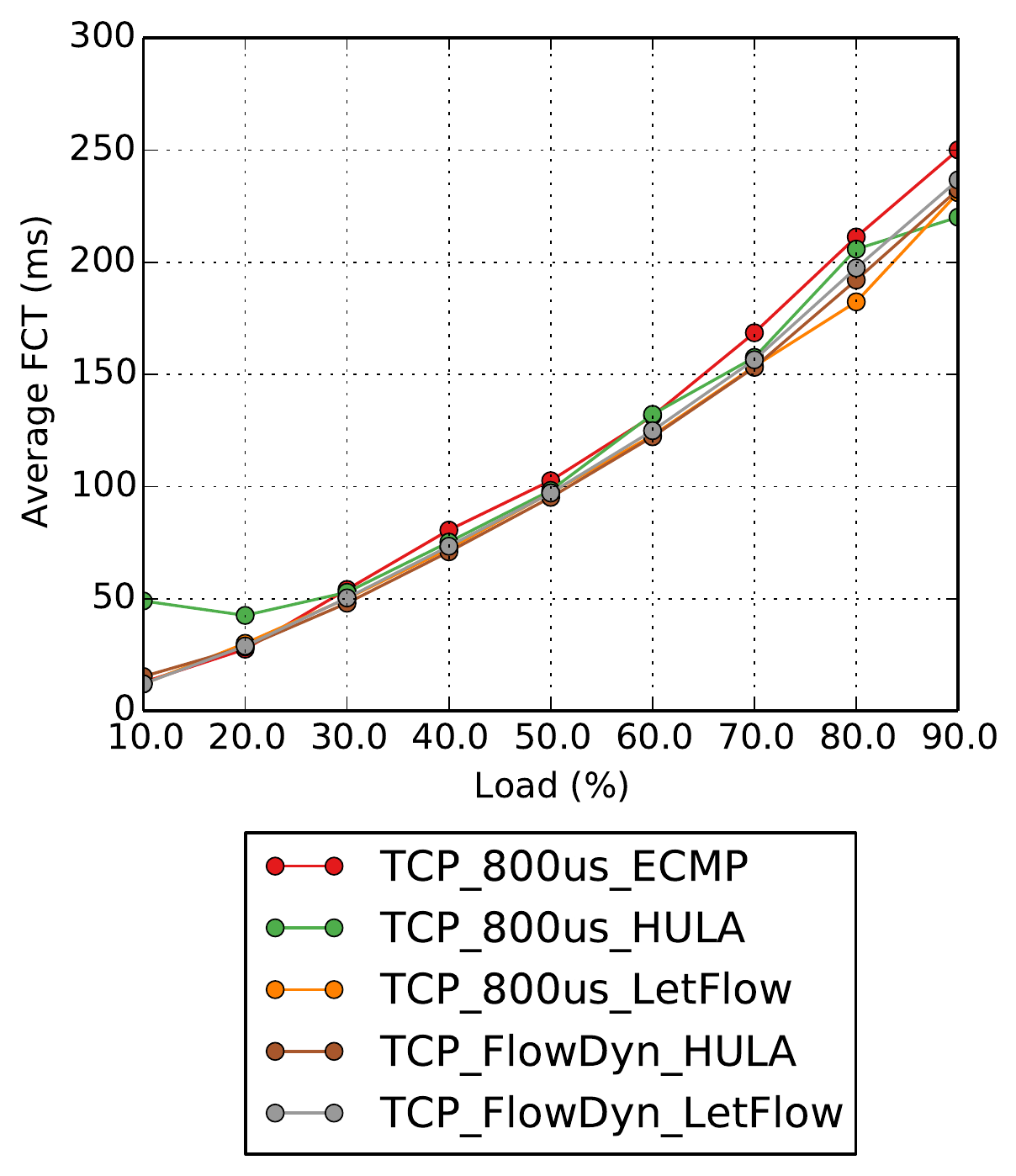}\par\caption{Average FCT using for data-mining traffic} \label{fig:Average_cdf_data}
   \end{minipage} \hfill
   \begin {minipage}{0.325\textwidth}
     \centering
     \includegraphics[width=.9\linewidth]{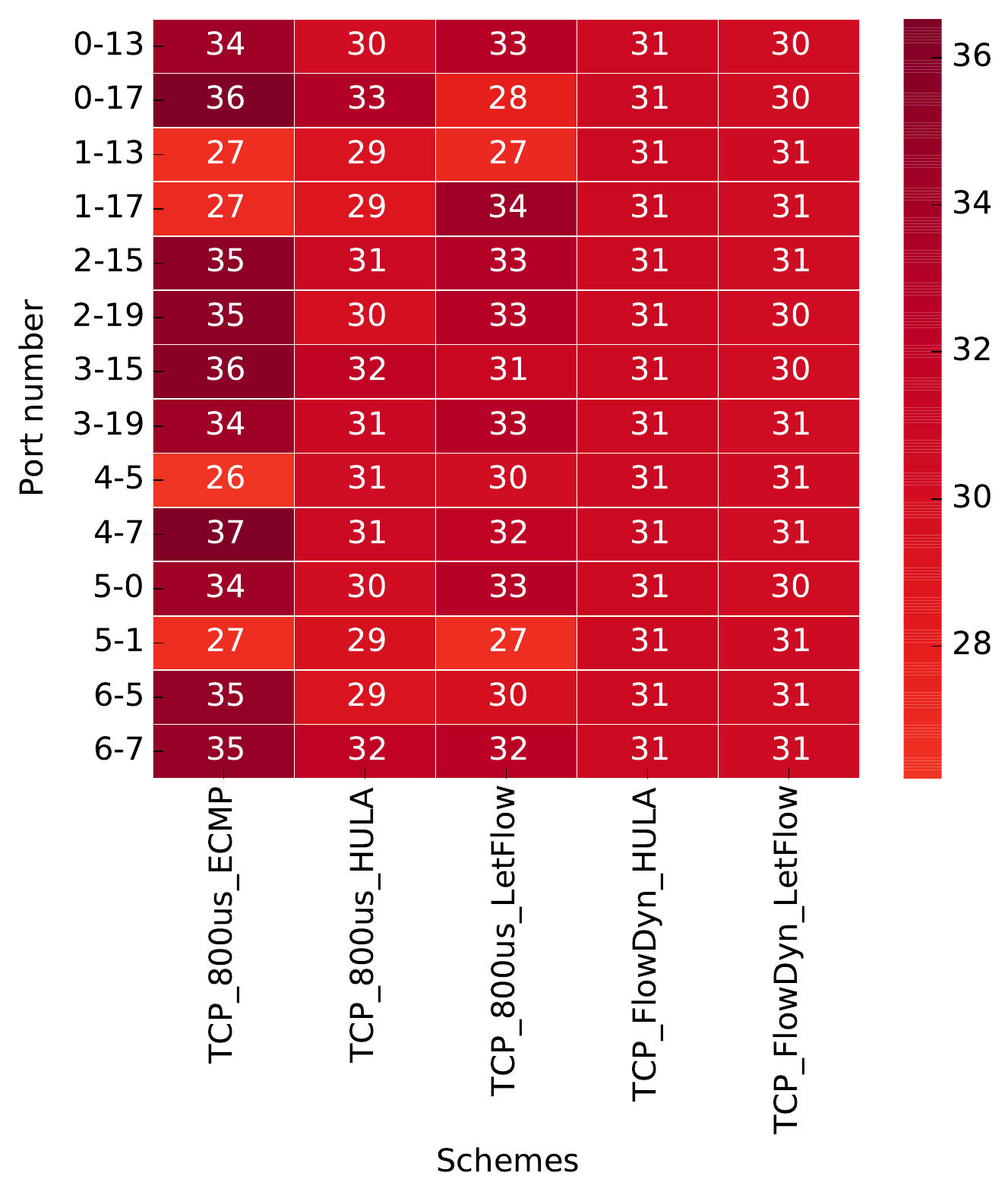}\par\caption{Link utilization (web-search)} \label{fig:Heat_map_cdf} 
   \end{minipage} \hfill
     \begin {minipage}{0.325\textwidth}
     \centering
	\includegraphics[width=.9\linewidth]{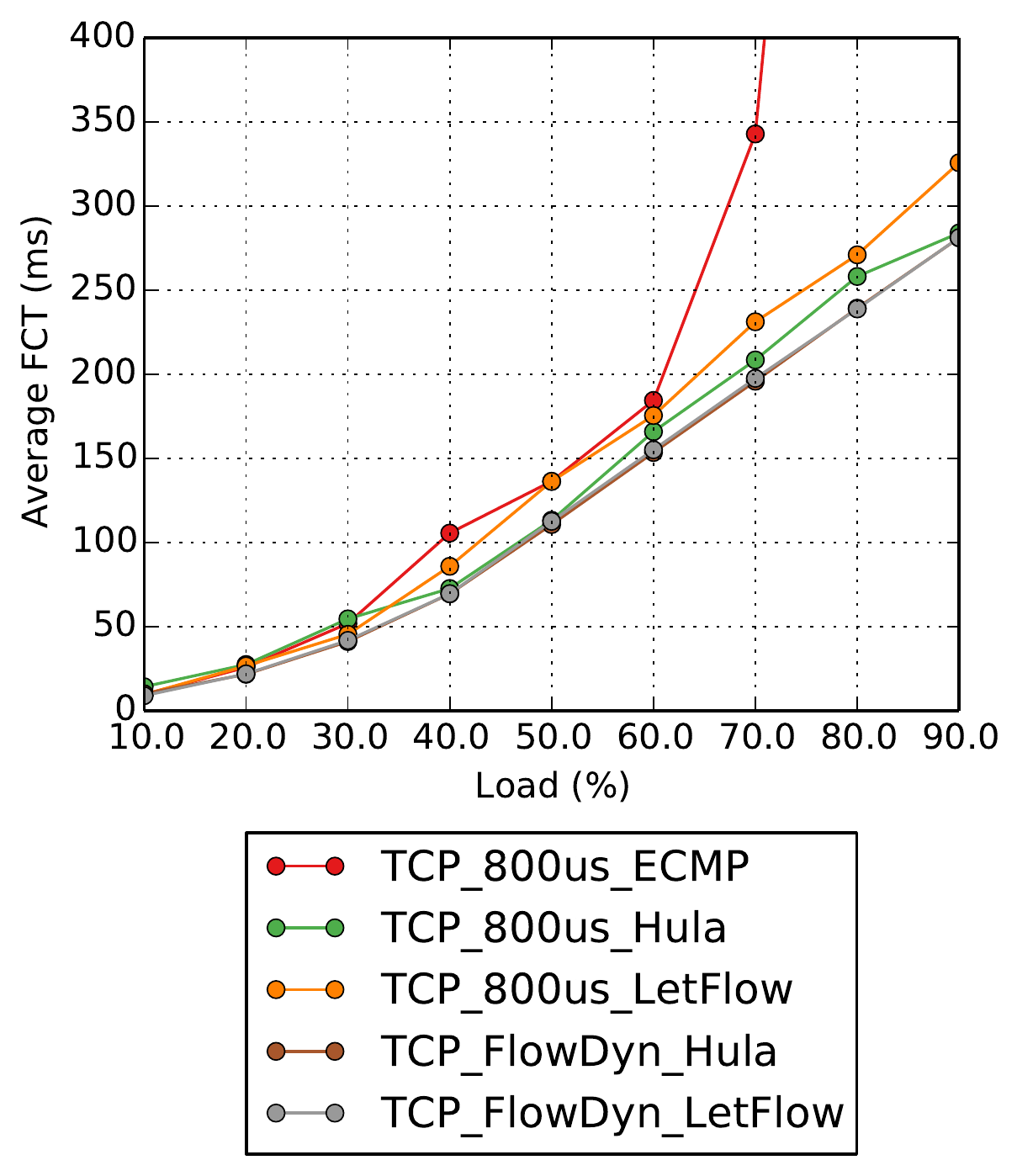}\par\caption{Average FCT for asymmetric topology with web-search traffic} \label{fig:Average_cdf_assymetric}
   \end{minipage}
\end{figure*}

\textbf{Schemes:} We compare ECMP as a baseline against HULA, and LetFlow. 
ECMP calculates the path of each flow  based on 5-tuple hash performed hop-by-hop. HULA and LetFlow are  flowlet-based load-balancing schemes. 
On the one hand, HULA uses probes to calculate the best next-hop to each destination ToR based on the maximum utilization calculated hop-by-hop using P4. 
On the other hand, LetFlow is congestion-agnostic and it is a simple approach that selects the next-hop for a flowlet randomly. 

\textbf{Workloads:} Each host generates traffic through a client-server application. We use traces from \cite{katta2016hula}, which emulate web-search and data-mining traffic. Each node initiates a TCP connection to a random node located in another POD. The amount of data sent is obtained by samples from the CDF of the selected workload trace and the inter-arrival time is modeled following an exponential Poisson process. We scale the traffic in a range from 10\% to 90\% by increasing the number of parallel connections from 1 to 9 respectively. All 32 nodes located in POD 1 and 2 act as clients, which select randomly any of the 32 nodes located in POD 3 and 4 that act as servers. Therefore, we will have from 32 parallel connections at 10\% of load, up to 288 parallel connections at 90\% load.

\textbf{Parameters:} We compare FlowDyn, which uses dynamic flowlet timeout, against the aforementioned load-balancing schemes that have the flowlet timeout set statically to 800$\mu$s, which is the maximum RTT of the network. The probing frequency is set to 100$\mu$s. The flowlet timeout intervals of FlowDyn is set in ranges of 100$\mu$s and a threshold of 70\%.

\subsection{Symmetric Topology}
First, we compare the different load-balancing schemes on the symmetric topology. Figure \ref{fig:Average_fct_dctcp} shows the average flow completion time for all flows for different network load. Regardless of the load-balancing scheme used, a dynamic flowlet timeout (FlowDyn) achieves a lower flow completion time (FCT) than when using a static flowlet timeout value. At higher traffic intensity, HULA with FlowDyn gets better results compared to ECMP and LetFlow since HULA leverage the use of probes to obtain congestion information of the network. On the contrary, at lower loads, if we configure HULA with a high flowlet timeout value to avoid re-ordering when latency increases, this causes HULA without FlowDyn to behave worse by losing many load-balancing opportunities. Figure \ref{fig:mice_websearch} shows separately the overall performance of web-search traffic for mice flows, i.e., those flows with a flow size less than 100KB. On the other hand, Figure \ref{fig:Average_elephant} illustrates for the same web-search traffic the average flow completion time for elephant flows, i.e., flows larger than 10MB.
Mice flows give fewer opportunities to balance the load since the vast majority of the flows have less than a hundred packets. In the case of elephant flows, load-balancing schemes have more opportunities to balance the load since TCP attempts to adapt to network congestion while receiving feedback in terms of packet loss or ACK packets. In both cases, FlowDyn achieves better performance, even though at 90\% of the load, the difference between any scheme with FlowDyn and without FlowDyn is reduced. This is because the static flowlet timeout is set to the optimum in high traffic conditions. Figure \ref{fig:Average_cdf_data} shows the average FCT for the data-mining workload. In this case, this traffic consists of 80\% of mice flows and long elephant flows. Therefore, at a low traffic intensity, the static flowlet timeout is too large and it misses more opportunities to balance the load compared to FlowDyn. 
FlowDyn with HULA achieves 3.19 times smaller FCT at 10\% network load compared with HULA without FlowDyn.

Figure \ref{fig:Heat_map_cdf} illustrates the maximum link utilization (Gbps) at 50\% load for different ports, illustrated in Figure \ref{fig:topo}. Since FlowDyn dynamically adjusts the flowlet timeout according to the network latency conditions, FlowDyn achieves a more uniform load balancing across all available paths. Therefore, it balances the load better by congesting all available paths more equally.

\subsection{Asymmetric Topology}
To simulate an asymmetric topology, we simulate a failure in core switch 0 by disabling the switch (Figure \ref{fig:topo}). As a result, the total available bandwidth is reduced by 25\%. Figure \ref{fig:Average_cdf_assymetric} shows the average FCT for web-search traffic with the asymmetric topology. Since ECMP and LetFlow are congestion-agnostic, they attempt to equally balance the traffic to both core switch 0, and 1, and therefore leading to switch 1 to become the bottleneck. LetFlow with FlowDyn can achieve 1.16 times smaller FCT at 90\% load compared to LetFlow with the static flowlet timeout value.

\section{Conclusions and Future Work}
\label{conclusion}
In this paper, we proposed FlowDyn, which uses programmable data planes to dynamically calculate a suitable flow and timeout flowlet that can be used in any flowlet-based load balancing such as HULA, CONGA and LetFlow. FlowDyn uses periodic probes to track latency differences between paths and adjusts the flowlet timeout according to the actual traffic load of the data center.

Our evaluation in a packet-level simulator shows that FlowDyn adapts to traffic changes and can generally improve the performance of load-balancing schemes in a large number of scenarios. 
FlowDyn can achieve 3.19 times smaller average flow completion time at 10\% network load and 1.16x at 90\% network load.
As our future work, we intend to compare the result obtained in the simulator with results obtained using real measurements from P4 programmable devices such as Barefoot Tofino, NetFPGAs, and SmartNics. In addition, our goal is not only to calculate the minimum flowlet timeout that allows guaranteeing the transport of burst of packets without introducing re-ordering, but also to find the optimal flowlet timeout.

\section*{Acknowledgment}
The authors would like to thank the reviewers for their comments that helped to improve the paper. Parts of this research have been funded by the Knowledge Foundation of Sweden through the HITS profile.

\def\bibfont{\footnotesize}
\bibliographystyle{IEEEtran}
\bibliography{example}
\end{document}